\documentclass[%
reprint,
superscriptaddress,
amsmath,
amssymb,
aps,
twocolumn,
showpacs,
pra,
]{revtex4-2}
\usepackage{tikz}
\usepackage{graphicx}
\usepackage{dcolumn}
\usepackage{bm}
\usepackage{braket}
\usepackage{hyperref}
\usepackage[caption=false]{subfig} 
\usepackage{tabularx}
\usepackage{booktabs}
\usepackage{amsmath}
\usepackage{graphicx}
\usepackage{fancyhdr}
\usepackage{float}
\usepackage{siunitx}
\usepackage{xcolor}
\usepackage{xspace}
\usepackage{eucal}
\usepackage{multirow}
\usepackage{lineno}
\usepackage{pgfplots}
\pgfplotsset{compat=1.14}
\newcommand{\Rb}{\textsuperscript{87}Rb\xspace}

\newcommand{\component}[1]{(#1)\xspace}

\newcommand{\unc}[2]{$#1(#2)$}

\newcommand{\uncexp}[3]{$#1(#2) \times 10^{#3}$}

\newcommand{\tempmin}{\unc{3.2}{0.6}~\si{\nano\kelvin}} 
\newcommand{\expmin}{\unc{553}{49}~\si{\micro\meter\per\second}}
\begin{document}
\title{
All-Optical Matter-Wave Lens using Time-Averaged Potentials
}
\author{H.~Albers}
\affiliation{Leibniz Universit\"at Hannover, Institut f\"ur Quantenoptik,\\ Welfengarten 1, 30167 Hannover, Germany}
\author{R.~Corgier}
\altaffiliation{
LNE-SYRTE, Observatoire de Paris, Université PSL, CNRS,
Sorbonne Université 61 avenue de l’Observatoire, 75014 Paris, France}
\affiliation{Leibniz Universit\"at Hannover, Institut f\"ur Quantenoptik,\\ Welfengarten 1, 30167 Hannover, Germany}
\affiliation{Universit\'e Paris-Saclay, CNRS, Institut des Sciences Mol\'eculaires d'Orsay, 91405 Orsay, France}
\author{A.~Herbst}
\author{A.~Rajagopalan}
\affiliation{Leibniz Universit\"at Hannover, Institut f\"ur Quantenoptik,\\ Welfengarten 1, 30167 Hannover, Germany}
\author{C.~Schubert}
\affiliation{Leibniz Universit\"at Hannover, Institut f\"ur Quantenoptik,\\ Welfengarten 1, 30167 Hannover, Germany}
\affiliation{Deutsches Zentrum für Luft- und Raumfahrt e.V. (DLR), Institut für Satellitengeod\"asie und Inertialsensorik, c/o Leibniz Universit\"at Hannover, DLR-SI, Callinstraße 36, 30167, Hannover, Germany}
\author{C.~Vogt}
\author{M.~Woltmann}
\author{C.~Lämmerzahl}
\author{S.~Herrmann}
\affiliation{ZARM Zentrum für angewandte Raumfahrttechnologie und Mikrogravitation, Universität Bremen,\\ Am Fallturm 2, 28359 Bremen, Germany}
\author{E.~Charron}
\affiliation{Universit\'e Paris-Saclay, CNRS, Institut des Sciences Mol\'eculaires d'Orsay, 91405 Orsay, France}
\author{W.~Ertmer}
\affiliation{Leibniz Universit\"at Hannover, Institut f\"ur Quantenoptik,\\ Welfengarten 1, 30167 Hannover, Germany}
\affiliation{Deutsches Zentrum für Luft- und Raumfahrt e.V. (DLR), Institut für Satellitengeod\"asie und Inertialsensorik, c/o Leibniz Universit\"at Hannover, DLR-SI, Callinstraße 36, 30167, Hannover, Germany}
\author{E.~M.~Rasel}
\author{N.~Gaaloul}
\author{D.~Schlippert}\email{Email: schlippert@iqo.uni-hannover.de}
\affiliation{Leibniz Universit\"at Hannover, Institut f\"ur Quantenoptik,\\ Welfengarten 1, 30167 Hannover, Germany}
\date{\today}
%
\begin{abstract}
The precision of matter-wave sensors benefits from interrogating large-particle-number atomic ensembles at high cycle rates.
Quantum-degenerate gases with their low effective temperatures allow for constraining systematic errors towards highest accuracy, but their production by evaporative cooling is costly with regard to both atom number and cycle rate.
In this work, we report on the creation of cold matter-waves using a crossed optical dipole trap and shaping them by means of an all-optical matter-wave lens.
We demonstrate the trade off between lowering the residual kinetic energy and increasing the atom number by reducing the duration of evaporative cooling and estimate the corresponding performance gain in matter-wave sensors.
Our method is implemented using time-averaged optical potentials and hence easily applicable in optical dipole trapping setups.
\end{abstract}
\maketitle
\section{Introduction}
Ever since their first realization, atom interferometers~\cite{Kasevich91PRL, Kasevich92APB,Riehle91PRL,Cronin09RMP} have become indispensable tools in fundamental physics~\cite{Biedermann15PRA, Bouchendira11PRL, Parker2018Science, Damour96CQG, Schlippert14PRL, Albers2020, Fray04PRL, Bonnin13PRA, Kuhn14NJP, Tarallo14PRL, Zhou15PRL,Asenbaum2020, Tino21IOP} and inertial sensing~\cite{Peters99Nature, Peters01Metrologia, Louchet11NJP,Freier16CS, Barrett2016, Hardman16PRL, Gersemann2020EPJD,Savoie2018SciAdv,Berg15PRL,Stockton11PRL,Gauguet09PRA,Canuel06PRL,Geiger2020}.
The sensitivity of such matter-wave sensors scales with the enclosed space-time area which depends on the momentum transferred by the beam splitters as well as the time the atoms spend in the interferometer.\\
The expansion of the atomic clouds, used in interferometers, needs to be minimized and well controlled to reach long pulse separation times, control systematic shifts, and create ensembles dense enough to detect them after long time-of-flights.
Nevertheless, colder ensembles with lower expansion rates typically need longer preparation times. 
Therefore, matter-wave sensors require sources with a high flux of large cold atomic ensembles to obtain fast repetition rates.\\
Bose-Einstein condensates (BECs) are well suited to perform interferometric measurements. 
They are investigated to control systematic effects related to residual motion at a level lower than a few parts in $10^9$ of Earth's gravitational acceleration~\cite{Hensel2021EPJD,Heine2020EPJD,Karcher2018NJP,Schkolnik15APB,Louchet11NJP}.
In addition, due to their narrower velocity distribution~\cite{Szigeti12NJP}, BECs offer higher beam splitting efficiencies and thus enhanced contrast~\cite{Abend16PRL,Hardman16PRL,Dickerson13PRL}, especially for large momentum transfer~\cite{Gebbe2021,Abend16PRL,Mcdonald13PRA,Chiow11PRL,Debs11PRA, Chiow09PRL,Clade09PRL}.
Finally, the inherent atomic collisions present in BECs can enhance matter-wave interferometry by enabling (i) ultra-low expansion rates through collective mode dynamics with a recent demonstration of a 3D expansion energy of $k_B\cdot38^{+6}_{-7}$~pK~\cite{Deppner2021PRL}, and (ii) ultimately the generation of mode entanglement through spin-squeezing dynamics to significantly surpass the standard-quantum limit~\cite{Kruse2016PRL,Szigeti2020,Anders2020arXiv,Corgier2021}.\\
Today's fastest BEC sources rely on atom-chip technology, where near-surface magnetic traps allow for rapid evaporation using radio frequency or microwave transitions.
This approach benefits from constant high trapping frequencies during the evaporative cooling process, thus leading to repetition rates on the order of \SI{1}{Hz} with BECs comprising $10^5$ atoms~\cite{Rudolph15NJP}.\\
Anyway, since magnetic traps are not suitable in certain situations optical dipole traps become the tool of choice~\cite{Chu86PRL}.
Examples are trapping of atomic species with low magnetic susceptibility~\cite{Stellmer2013PRA,Roy2016PRA}, or molecules~\cite{Anderegg2018, Carr2009} and composite particles~\cite{Grimm00AAMOP,Martinez2016}.
In optical dipole traps external magnetic field allow tuning parameters, e.g., when using Feshbach resonances~\cite{Salomon14PRA}.\\
But the intrinsic link between trap depth and trap frequencies in dipole traps~\cite{OHara01PRA} inhibits runaway evaporation.
Cold ensembles can be only produced in shallow traps, leading to drastically increased preparation time $t_\text{P}$.
This long standing problem has been recently overcome through the use of time-averaged potentials, where trap depth and trap frequencies can be controlled independently, thus allowing for more efficient and faster evaporation while maintaining high atom numbers~\cite{Roy2016PRA,Condon2019}.\\
In this work, we use dynamic time-averaged potentials for efficient BEC generation and demonstrate an all-optical matter-wave lens capable of further reducing the ensemble's residual kinetic energy.
Contrary to pulsed schemes of matter-wave lensing~\cite{Ammann97PRL, Morinaga99PRL, Myrskog2000, Luan2018, KanthakNJP2021, Gochnauer2021Atoms, Deppner2021PRL}, we keep the atoms trapped over the entire duration of the matter-wave lens~\cite{Dickerson13PRL}, which eases implementation in ground-based sensors.
Moreover we show that with this technique one can short-cut the evaporation sequence prior to the matter-wave lens,
which increases the atomic flux by enhancing atom number and reducing cycle time while simultaneously reducing the effective temperature.
Our method can largely improve the matter-wave sensor's stability in various application scenarios.
\section{Results}\label{sec:results}
\subsection{Evaporative cooling}
We operate a crossed optical dipole trap at a wavelength of \SI{1960}{\nano\meter} loaded from a \Rb magneto-optical trap~(details in the ``Methods'' section).
The time-averaged potentials are generated by simultaneuos center-position modulation of the crossed laser beams in the horizontal plane.
Controlling the amplitude of this modulation and the intensity of the trapping beams enables the dynamic control and decoupling of the trapping frequencies and depth.
We chose the waveform of the center-position modulation that generates a parabolic potential~\cite{Roy2016PRA}. \\
\begin{figure}
    \begin{center}
    \resizebox{1\columnwidth}{!}{\includegraphics{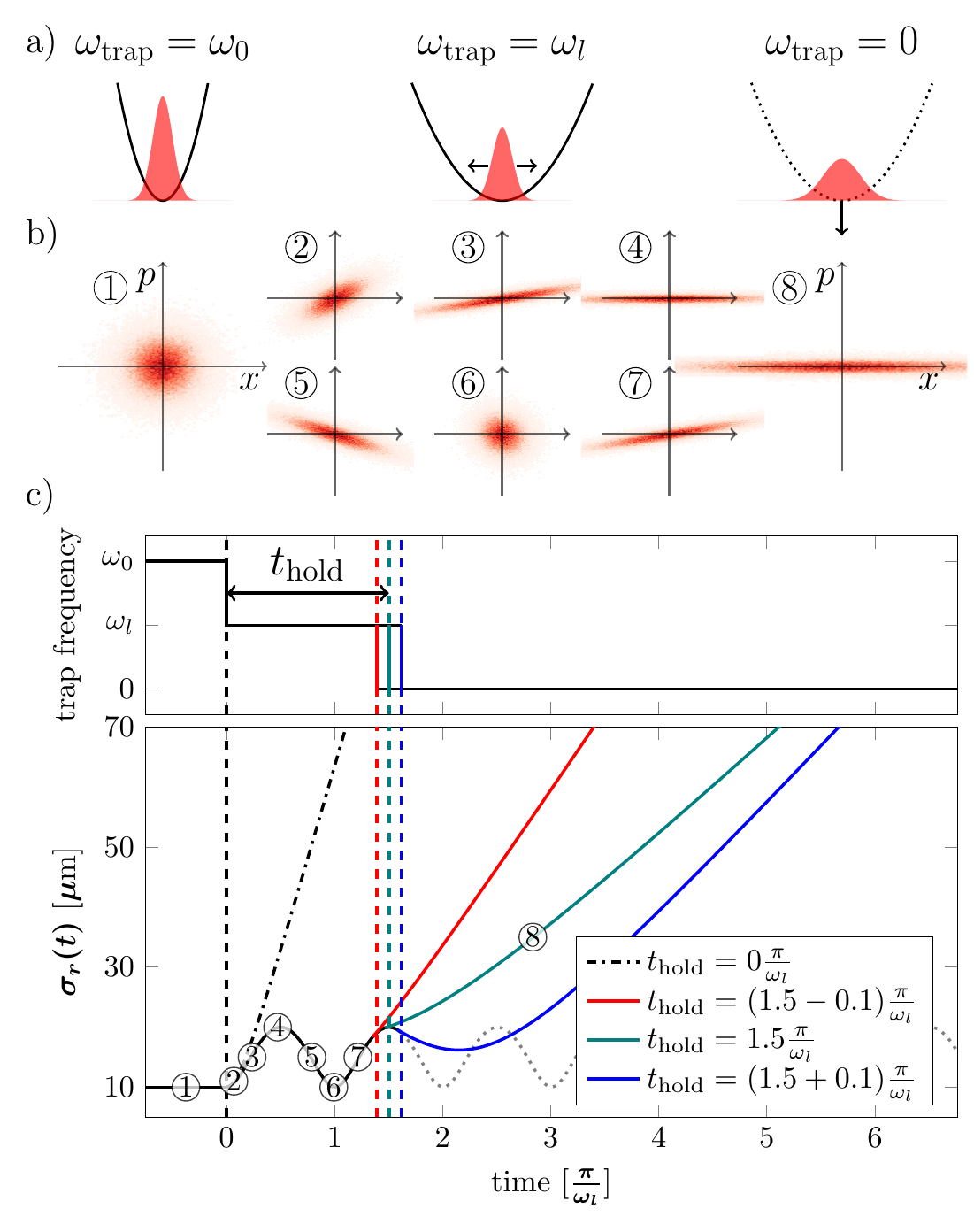}}
    \caption{\textbf{Scheme of the matter-wave lens.} 
    The drawing in a) shows the three trap configurations and the distribution of the atomic ensemble during the matter-wave lens.
    The phase-space diagrams in b) show the atomic distributions at different timings during the matter-wave lens marked with numbers in b) and c).
    The upper graph in c) shows the behavior of the time dependency of the trapping frequency, while the lower graph shows the simulated evolution of the size of the atomic ensemble.
    After the holding time in the initial trap the trapping frequency is rapidly decreased at time $t=$~\SI{0}{ms}. 
    The size of the atomic ensemble starts to oscillate (solid and dotted black line).
    At time $t_{\text{hold}}=(n+0.5)\times \pi/\omega_l$, with $n \in \mathbb{N}$, this oscillation reaches an upper turning point (teal curve).
    The atomic ensemble is released at its maximum size to minimize its later expansion rate.
    If the release time ($t_{\text{hold}}$) does not match this condition the expansion rate is not minimized (red and blue curve).
    The dashed-dotted black curve displays the size of a free falling ensemble without lensing, released at time $t=$~\SI{0}{ms}.}
    \label{fig:MWL_size_sim}
    \end{center}
\end{figure}
\begin{figure}
    \begin{center}
    \resizebox{1\columnwidth}{!}{\includegraphics{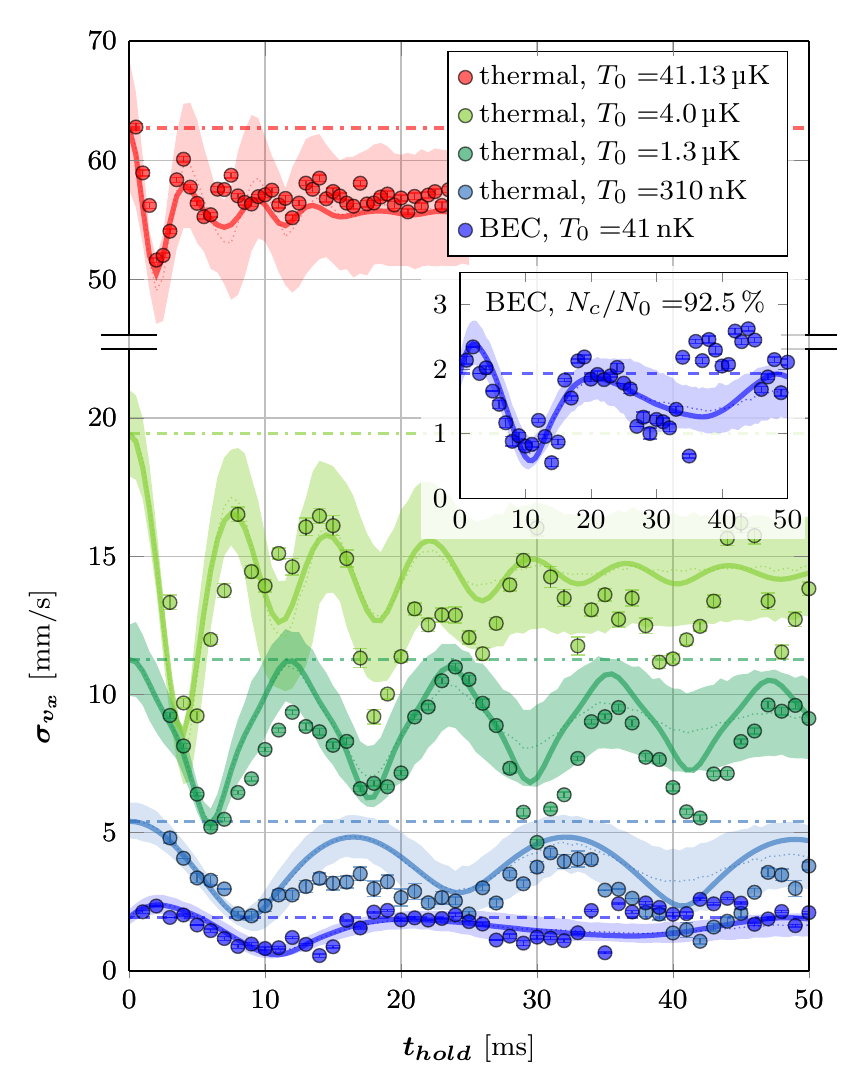}}
    \caption{\textbf{Oscillations of expansion velocity.}
    Expansion velocity after \SI{30}{ms} of time of flight for different initial temperatures $T_0$ in dependence of the holding time.
    The circles show the measurements, the dashed-dotted lines the expansion rate from the initial traps~(cf. black dashed-dotted line in Fig.~\ref{fig:MWL_size_sim}). 
    The simulations use the scaling ansatz and are depicted as lines with a shaded 1-sigma error estimation for the used trap parameters.}
    \label{img:osci}
    \end{center}
\end{figure}
Up to \SI{2e7} rubidium atoms are loaded into the trap with trapping frequencies $\omega/2\pi \approx \{140;\, 200;\, 780\}$~\si{\hertz} in $\{x^\prime;\, y^\prime;\, z\}$ direction with a trap depth of \SI{170}{\micro\kelvin}.
For this we operate the trap at the maximum achievable laser intensity of \SI{12}{\watt} and the center-position modulation amplitude of $h_0=$~\SI{140}{\micro\meter}.\\
We perform evaporative cooling by reducing the trap depth exponentially in time while keeping the trapping frequencies at a high level by reducing the amplitude of the center-position modulation.
This method allows to generate BECs with up to \SI{4e5}{atoms} within \SI{5}{s} of evaporative cooling.
By shortening the time constant of the exponential reduction we generate BECs with \num{5e4} (\num{2e5})~\si{particles} within \SI{2}{\second} (\SI{3}{\second}) of evaporative cooling.
At the end of the evaporation sequence the trap has frequencies of $\omega/2\pi \approx \{105;\, 140;\, 160\}$\,\si{\hertz} and a depth of about \SI{200}{\nano\kelvin}.
The expansion velocity of the condensate released from the final evaporation trap is \SI{2}{\milli\meter\per\second}, which corresponds to an effective temperature of \SI{40}{\nano\kelvin}.
\subsection{All-optical matter-wave lens}
\begin{figure}
    \begin{center}
    \resizebox{1\columnwidth}{!}{\includegraphics{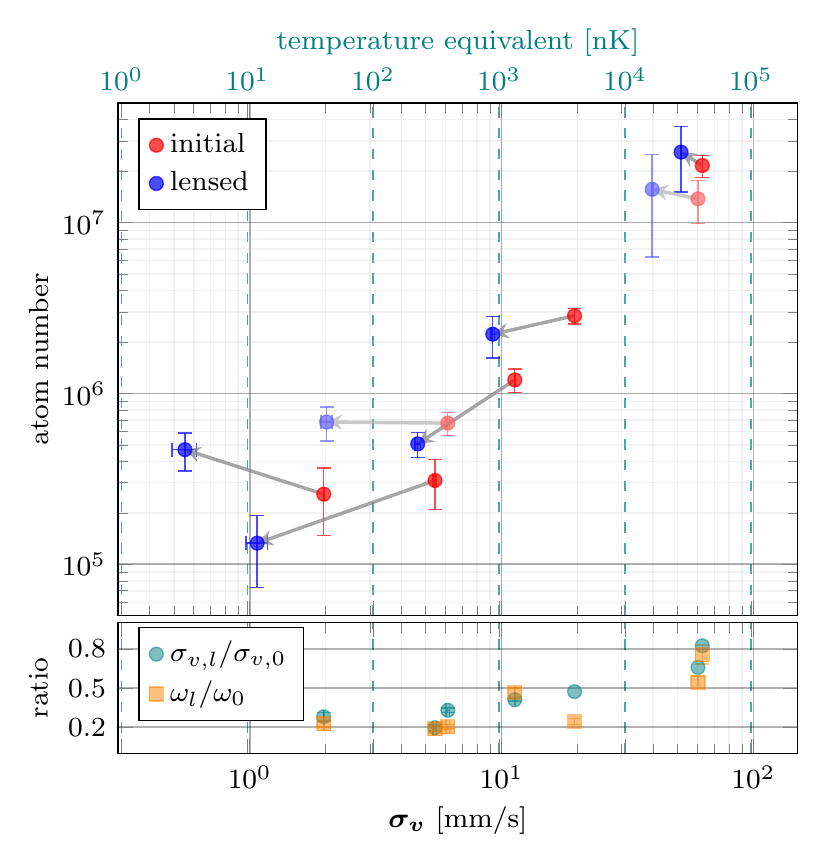}}
    \caption{\textbf{Expansion velocity dependent atom number.}
    Resulting expansion velocity (top graph, blue circles) after the matter-wave lens for different starting temperatures (red circles). 
    The lines connect the corresponding data points.
    The two grayed out data points associated with the starting expansion velocities \SI{60}{\milli\meter\per\second} and \SI{6}{\milli\meter\per\second} are not displayed in Fig.~\ref{img:osci} while all the others are included in Fig.~\ref{img:osci}.
    The bottom graph shows the ratio of the lensed and the initial ensembles expansion as well as the trapping frequencies.
    } 
    \label{img:flux_data}
    \end{center}
\end{figure}

Our matter-wave lens can be applied in any temperature regime explorable in our optical trap.
We investigate the creation of collimated atomic ensembles for different initial temperatures of the matter-waves. 
To this aim, the evaporation sequence is stopped prematurely at different times to generate input atomic ensembles at rest with initial trap frequency $\omega_0$ and initial temperature $T_0$.
We then initiate the matter-wave lens by a rapid decompression \cite{Chu86OL} of the trap frequency in the horizontal directions from $\omega_0$ to $\omega_l$.
Here we denote by $\omega_l$ to the lensing potential in analogy with the Delta-kick collimation technique. 
The reduction of the trapping frequencies from the initial $\omega_0$ to $\omega_l$ depends on experimental feasibility.
The ratio of $\omega_l/\omega_0$ for each measurement is shown in the bottom graph of Fig.~\ref{img:flux_data}.
It depends on the maximum achievable amplitude of the center-position modulation and the modulation amplitude right before the rapid decompression.
With ongoing evaporative cooling this amplitude is reduced and thus the trap can be relaxed much further for more continued sequence.
However, we need to maintain the confinement in the vertical direction by adjusting the dipole trap's intensity to suppress heating or loss of atoms.\\
Subsequent oscillations in the trap result in a manipulation in phase space (Fig.~\ref{fig:MWL_size_sim}~a) \& b)) for focusing, diffusion, and, importantly collimation of the matter-wave (Fig.~\ref{fig:MWL_size_sim}~c)).
Fig.~\ref{fig:MWL_size_sim}~c) depicts the expansion of a thermal ensemble in 1D for three different holding times ($t_{\text{hold}}$) to highlight the importance of a well chosen timing for the lens.
Fig.~\ref{img:osci} shows exemplary expansion velocities (colored circles) depending on the holding time $t_\text{hold}$.
The colored curves in this graph display the simulated behavior following the scaling ansatz~(details in the ``Methods'' section) with an error estimation displayed by shaded areas.
Only for the final measurement (also shown in the inset in Fig.~\ref{img:osci}) we create a BEC with a condensed fraction of \SI{92.5}{\percent} of the total atom number and apply the matter-wave lens to it.\\
With the presented method we observe oscillations of the expansion rate, which are in good agreement with the simulations for different ensemble temperatures.
For all investigated temperatures an optimal holding time exists for which the final expansion rate is minimized (Fig.~\ref{img:flux_data}).\\
The change in atom number from the initial to the lensing trap (Fig.~\ref{img:flux_data}) lies within the error bars and arises mainly due to pointing instabilities of the crossed optical dipole trap beams.
The lowest expansion rate is achieved with \expmin~with a related effective temperature of \tempmin~and an atom number of \uncexp{4.24}{0.02}{5}.
With this method we achieve a more than one order of magnitude lower effective temperature while maintaining a comparable atom number compared to evaporative cooling.
\section{Discussion}\label{sec:discussion}
In this paper we demonstrate a technique to reduce the expansion velocity of an atomic ensemble by rapid decompression and subsequent release from an dipole trap at a well-controlled time.
The efficiency of the matter-wave lens for higher temperatures is mainly limited experimentally by the limited ratio between the initial and the lensing trap frequency $\omega_{l}/\omega_{0}$ (Fig. 3, lower graph) which is constrained by the maximum possible spatial modulation amplitude of the trapping beams. 
In general, according to the Liouville theorem, the expansion speed reduction of the matter-wave is proportional to $(\omega_0/\omega_l)^2$ where a large aspect ratio enables a better collimated ensemble.
The atoms are loaded into the time-average potential with an optimized center-position modulation amplitude of \SI{140}{\micro\meter}, while the maximum is \SI{200}{\micro\meter}.
During the evaporation sequence this amplitude is decreased.
Consequently, the relaxation of the trap is less efficient at the beginning of the evaporative sequence or directly after the loading of the trap.\\
Another constraint is that the trap's confinement in the unpainted vertical direction is required to remain constant.
If the vertical trap frequency is increased we observe heating effects and suffer from atom loss when it is decreased.
To compensate for the trap depth reduction during the switch from the initial to the lensing trap we increase the dipole trap laser's intensity accordingly.\\
An additional modulation in the vertical direction, e.g., by means of a two-dimensional acousto-optical deflector, as well as an intersection angle of \SI{90}{\degree} would enable the generation of isotropic traps.
In such a configuration, the determination of the optimal holding time will benefit from the in-phase oscillations of the atomic ensemble's size~\cite{Li2019}.
When applying our matter-wave lens in a dual-species experiment, isotropy of the trap will also improve the miscibility of the two ensembles~\cite{Corgier2020NJP}.\\
To illustrate the relevance for atom interferometers, we discuss the impact of our source in different regimes~(details in the ``Methods'' section) operated at the standard quantum limit for an acceleration measurement.
In a Mach-Zehnder-like atom interferometer~\cite{Peters99Nature,Kasevich91PRL}, the instability reads
\begin{equation}\label{eq:instability}
	\sigma_a(\tau) = \frac{1}{C\sqrt{N}n k_{\text{eff}} T_{\text{I}}^2}\cdot \sqrt{\frac{t_{\text{cycle}}}{\tau}}
\end{equation}
after an averaging time $\tau$, neglecting the impact of finite pulse durations on the scale factor~\cite{Bertoldi2019PRA,Antoine07PRA,Cheinet08IEEE}.
Eq.~(\ref{eq:instability}) scales with the interferometer contrast $C$, the atom number per cycle $N$, the effective wave number $n\hbar k_{\text{eff}}$ indicating a momentum transfer during the atom-light interaction corresponding to $2n$ photons, and the separation time between the interferometer light pulses $T_{\text{I}}$. 
The cycle time of the experiment $t_\text{cycle} = t_{\text{P}} + 2T_{\text{I}} + t_{\text{D}}$ includes the time for preparing the ensemble $t_{\text{P}}$, the interferometer $2T_{\text{I}}$, and the detection $t_{\text{D}}$.
In Eq.~(\ref{eq:instability}), the contrast depends on the beam splitting efficiency.
This, in turn, is affected by the velocity acceptance and intensity profile of the beam splitting light, both implying inhomogeneous Rabi frequencies, and consequently a reduced mean excitation efficiency~\cite{Loriani2019NJP,Szigeti12NJP,Kasevich91PRL2}.
\begin{figure}
    \begin{center}
    \resizebox{1\columnwidth}{!}{\includegraphics{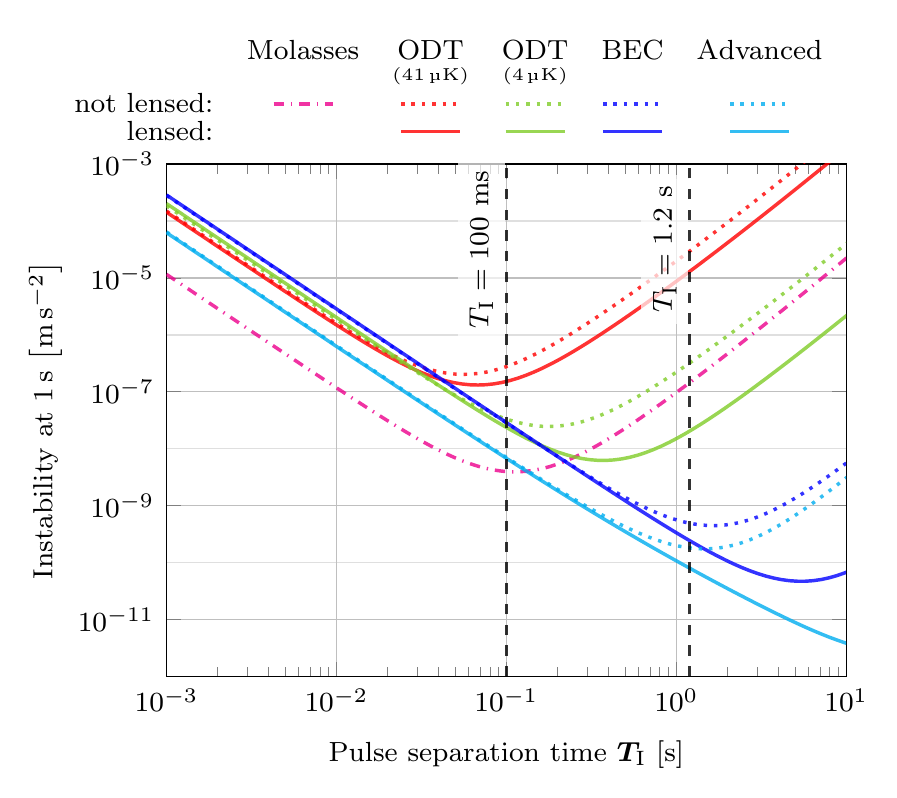}}
    \caption{\textbf{Instability comparison.}
    Behavior of the instability for shot-noise limited atom interferometers after an integration time of $\tau=$~\SI{1}{\second} for different sources, see Tab.~\ref{tab:instability}, over the pulse separation time $T_{\text{I}}$.
    The colors of the curves, except for the molasses and advanced case, agree with the measurements displayed in Fig.~\ref{img:osci}.
    The different source parameters are described in more detail in the ``Methods'' section.
    } 
    \label{img:instability}
    \end{center}
\end{figure}
Due to expansion of the atomic ensemble and inhomogeneous excitation, a constrained beam diameter implicitly leads to a dependency of the contrast $C$ on the pulse separation time $T_{\text{I}}$, which we chose as a boundary for our discussion.
We keep the effective wave-number fixed and evaluate $\sigma_ a(1\,\mathrm{s})$ for different source parameters when varying $T_{\text{I}}$.\\
Fig.~\ref{img:instability} shows the result for collimated (solid lines) and uncollimated (dotted lines) ensembles in our model (see sec.~\ref{sec:instability}) and compares them to the instability under use of a molasses-cooled ensemble~(dash-dotted line).
Up to $T_{\text{I}}=100\,\mathrm{ms}$ and $\sigma_a(1\,\mathrm{s})=10^{-8}\,\mathrm{m\,s}^{-2}$, the molasses outperforms evaporatively cooled atoms or BECs due the duration of the evaporation adding to the cycle time and associated losses.
In this time regime, the latter can still be beneficial for implementing large momentum transfer beam splitters~\cite{Gebbe2021,Abend16PRL,Mcdonald13PRA,Chiow11PRL,Chiow09PRL,Clade09PRL} reducing $\sigma_a(\tau)$ or suppressing systematic errors~\cite{Abe2021QST,Hensel2021EPJD,Heine2020EPJD,Karcher2018NJP,Schkolnik15APB,Louchet11NJP} which is not represented in our model and beyond the scope of this paper.
According to the curves, exploiting higher $T_{\text{I}}$ for increased performance requires evaporatively cooled atoms or BECs.
This shows the relevance for experiments on large baselines~\cite{Abe2021QST,Badurina2020JCAP,Hardman16PRL,Hartwig15NJP,Dickerson13PRL,Zhou11GRG} or in microgravity~\cite{Kulas2016, Vogt2019}.
We highlight the extrapolation for the Very Long Baseline Atom Interferometer (VLBAI)~\cite{Hartwig15NJP, Schlippert2020CPTproceed}, targeting a pulse separation time of $T_\text{I}=$~\SI{1.2}{\second}~\cite{Schilling2020JG}.
Here, the model describing our source gives the perspective of reaching picokelvin expansion temperatures of matter-wave lensed large atomic ensembles.\\
\section{Methods}
\subsection{Experimental Realization}\label{sec:experiment}
The experimental apparatus is designed to operate simultaneous atom interferometers using rubidium and potassium and is described in detail in references~\cite{Zaiser11PRA,Albers2020,Schlippert14PRL}.\\
\begin{figure}
    \begin{center}
    \resizebox{1\columnwidth}{!}{\includegraphics{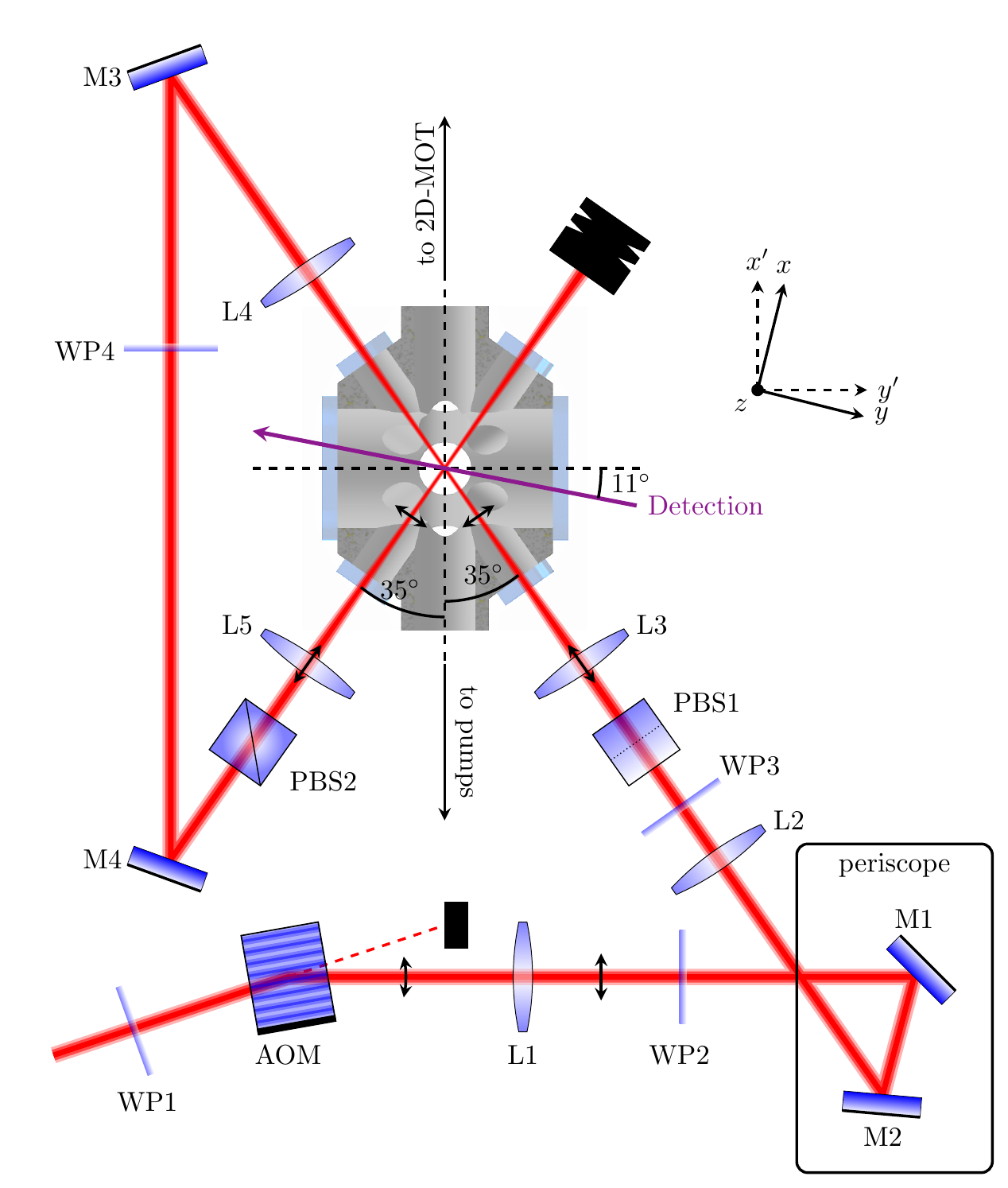}}
    \caption{\textbf{Experimental setup.}
    Optical setup of the dipole trap and alignment through the vacuum chamber. 
    The acousto-optical modulator (AOM) is used for modulating the center-position of the laser beams and intensity control.
    The $\lambda/2$ wave plate WP1 rotates the polarization of the beam for best diffraction efficiency.
    Lenses L1 ($f_1=\SI{100}{\milli\meter}$) and L2 ($f_2=\SI{300}{\milli\meter}$) magnify the beam radius to about \SI{3}{\milli\meter}, \SI{4}{\milli\meter} (vertical, horizontal).
    Downstream lenses L3, L4, and L5 ($f_{3,4,5}=\SI{150}{\milli\meter}$) focus, re-collimate and re-focus the beam into the center of the chamber.
    The $\lambda/2$ wave plates WP3, WP4, and the $\lambda/4$ wave plate WP2 set the polarization for maximum transmission at the orthogonal oriented polarization beam splitters (PBS1 and PBS2).
    The mirrors M1 and M2 form a periscope to guide the beam onto the level of the atoms, while M3 and M4 direct the beam a second time through the chamber.
    The purple arrow indicates the direction of absorption detection along the y-direction.}
    \label{fig:odt_setup}
    \end{center}
\end{figure}
For the experiments presented in this article only rubidium atoms were loaded from a two dimensional to a three dimensional magneto-optical trap (2D/3D-MOT) situated in our main chamber.
After \SI{2}{s} we turn off the 2D-MOT and compress the atomic ensemble by ramping up the magnetic field gradient as well as the detuning of the cooling laser in the 3D-MOT.
Subsequent to compression, the atoms are loaded into the crossed dipole trap by switching off the magnetic fields and increasing the detuning of the cooling laser to about \num{-30}~$\Gamma$, with $\Gamma$ being the natural linewidth of the $\text{D}_2$ transition.\\
Fig.~\ref{fig:odt_setup} depicts the setup of our crossed optical dipole trap.
The center-position modulation of the trapping beams is achieved by modulating the frequency driving the acousto-optical modulator (AOM) \component{Polytec, ATM-1002FA53.24}.
A voltage-controlled oscillator \component{Mini-Circuits, ZOS-150+} generates the signal for this, which is driven by a programmable arbitrary-waveform generator \component{Rigol, DG1022Z}.
We chose the waveform to generate a large-volume parabolic potential based on the derivation shown in Ref.~\cite{Roy2016PRA}. 
The amplitude of the displacement of the center-position of the dipole trap beam, $h_0$, is controlled by regulating the amplitude of the AOM's frequency modulation.
This yields a maximum beam displacement of $h_0=\SI{200}{\micro\meter}\, (\SI{300}{\micro\meter})$ at the position of the atoms for the initial (recycled) beam.
\subsection{Data acquisition and analysis}\label{sec:measurement}
We apply our matter-wave lens subsequent to loading the dipole trap and evaporative cooling.
The duration of the complete evaporative sequence is \SI{5}{s} for the measurements presented here we interrupt this sequence after \SIlist{0;0.2;1;2;3.5;4.3;5}{\second}.
Before the step-wise change of the trap frequency ($\omega_0 \rightarrow \omega_l$) we hold the ensemble in the trap given by the respective evaporation step configuration for \SI{50}{\milli\second}.\\
During the matter-wave lens, the rapid decompression of the trap causes oscillations of the ensemble's radius in the lensing trap.
Depending on the release time we observe oscillations by performing absorption imaging with iterating $t_{\text{hold}}$ for different times after the release from the trap.
For each holding time the expansion velocity is extracted by fitting a ballistic expansion.
This expansion can be transformed into an effective temperature using:
\begin{equation}\label{eq:sigma_v}
    \sigma_{v_i}^2 = \frac{k_B T_i}{m} \qquad ,
\end{equation}
along each direction.
The measurement is performed for different starting temperatures in the thermal regime as well as the BEC.\\
The simulations shown in Fig.~\ref{img:osci} use the scaling ansatz as described in Sec.~\ref{sec:scaling_approach}.
Here, the trapping frequencies of the lens potential in $x$- and $y$-direction have been extracted by fitting two damped oscillations to the measured data.
The starting expansion velocity was set by choosing a reasonable initial radius of the ensemble~(Table~\ref{tab:instability}).
The other parameters arise from the measurements or simulations of the trapping potentials.
The shaded areas in figure~\ref{img:osci} depict an error estimation of the expansion velocity oscillations obtained from performing the simulation by randomly choosing input parameters from within the error bars for 1000 simulation runs and calculating the mean value as well as the standard deviation for each $t_\text{hold}$.
\subsection{Scaling Ansatz}\label{sec:scaling_approach}
In the case of a thermal ensemble in the collision-less regime, the dynamics of a classical gas can be described using the scaling ansatz of Refs.~\cite{Guery-Odelin2002PRA,Pedri2003PRA}, which we briefly recall here for sake of simplicity. 
Here, the size of the ensemble scales with the time dependent dimensionless factor $b_i(t)$.
\begin{eqnarray}
	&\ddot{b_i}(t) + \omega_{i}^2(t) b_i(t) - \omega_{0,i}^2\dfrac{\theta_i(t)}{b_i(t)} \nonumber\\
	&+ \omega_{0,i}^2 \xi\left( \dfrac{\theta_i(t)}{b_i(t)} - \dfrac{1}{b_i(t) \prod_j b_j(t)}\right)  = 0 \label{eq:scaling}\\
	&\dot{\theta_i}(t) + 2 \dfrac{\dot{b_i}(t)}{b_i(t)}\theta_i(t) + \dfrac{1}{\tau}\left(\theta_i(t) - \frac{1}{3}\sum_j \theta_j(t)\right) = 0
	\quad ,
    \label{eq:effective temp scaling}
\end{eqnarray}
where $\theta_i$ acts as an effective temperature in the directions $i \in {x,y,z}$.
Here $\omega_{0,i}$ denotes the initial angular trap frequency and $\omega_{i}(t)$ denotes the time-dependent angular trap frequency defined such as: $\omega_{i}(t)=\omega_{l,i}$ for $0<t<t_\text{hold}$, with $\omega_{l,i}$ being the lensing potential, and $\omega_{i}(t)=0$ after the release (see Fig.\ref{fig:MWL_size_sim}).
This system of coupled differential equations contains the mean field interaction, given by the factor:
\begin{equation}\label{eq:E mean field}
	\xi = \dfrac{E_{mf}}{E_{mf} + k_B T} \qquad ,
\end{equation}
with 
\begin{equation}
	E_{mf} = \dfrac{4 \pi \hbar^2 a_s n_0}{m} \qquad ,
\end{equation}
were $a_s$ is the s-wave scattering length, $n_0$ the peak density and m the mass of a single particle.
Collision effects are also taken into account through
\begin{equation}\label{eq:relexation time}
	\tau = \tau_0 \times \left(\prod_j b_j\right) \times \left(\frac{1}{3}\sum_k \theta_k\right)
\end{equation}
with the relaxation time
\begin{equation}
	\tau_0 = \dfrac{5}{4 \gamma}
\end{equation}
and~\cite{Pedri2003PRA}
\begin{equation}
	\gamma = \frac{2}{\sqrt{2 \pi}} n_0 \sigma_{coll} \sqrt{\dfrac{k_B T}{m}} \qquad .
\end{equation}
In the special case of a BEC, the mean field energy is large compared to the thermal ensemble’s energy ($\xi \approx 1$) and the time scale on which collisions appear goes to zero ($\tau \approx 0$). 
In this case the time dependent evolution of the matter-wave can be described following Ref.~\cite{Castin96PRL}. 
Here, the evolution of the BEC's Thomas-Fermi radius, $R_i(t) = b_i(t) R_i(0)$, is described by the time-dependent evolution of the scaling parameter: 
\begin{equation}
    \ddot{b}_i(t) + \omega_i^2(t) b_i(t) = \dfrac{\omega_i(0)}{b_i(t) b_x(t) b_y(t) b_z(t)}
\end{equation}
and $R_i(0)$ is the initial Thomas-Fermi radius of the BEC along the i-th direction. 
It is worth to notice that recent studies \cite{Modugno2018, Viedma2020} extend the analysis of Refs.~\cite{Guery-Odelin2002PRA,Pedri2003PRA} to the BEC regime described in Ref.~\cite{Castin96PRL}.\\
With this set of equations the time evolution of the ensemble's size ($\sigma_{r_i}$) and velocity distribution ($\sigma_{v_i}$) is determined during the entire sequence of our matter-wave lensing sequence by 
\begin{equation}\label{eq:size evolution}
	\sigma_{r_i}(t) = \sigma_{r_i}(0) \times b_i(t)
\end{equation}
and
\begin{equation}\label{eq:velocity spread evolution}
	\sigma_{v_i}(t) = \frac{d \sigma_{r_i}(t)}{dt} \qquad .
\end{equation}
The scaling parameter $b_i$ can be applied either on the radius of a gaussian distributed thermal ensemble or the Thomas-Fermi radius of a BEC.
\subsection{Estimation of instability in matter-wave sensors}\label{sec:instability}
The instability of a matter-wave sensor operating at the standard quantum limit can be estimated using Eq.~(\ref{eq:instability}).
Here we assume Raman beam splitters ($n=1$) with a $1/e^2$-radius of \SI{1.2}{\centi\meter} and a pulse duration of $t_\pi=$~\SI{15}{\micro\second}.
The contrast ($C$) is taken into account as the product of the excitation probabilities of the atom-light interactions during the Mach-Zehnder type interferometer following \cite{Loriani2019NJP}.
Table~\ref{tab:instability} shows the source parameters used for the estimation of the instability.
\begin{table}
    \caption{\textbf{Source parameters for the instability estimation} for molasses cooled, released from the optical dipole trap (ODT) with and without evaporation, and condensed ensembles. 
    The values for the ensemble radius ($\sigma_r$) and expansion velocity ($\sigma_v$) are given for the horizontal (h) and vertical (v) direction, which corresponds to the transverse and longitudinal direction of the beam splitter respectively.}
    \begin{tabular}{@{}lr@{\hskip 0mm}lr@{\hskip 0mm}lrc@{}}
    	\toprule
    	\toprule
    	 & \multicolumn{2}{c}{$\sigma_r$ (h/v)} & \multicolumn{2}{c}{$\sigma_v$ (h/v)} & \multicolumn{1}{c}{$N$} & $t_\text{P}$ \\
    	 & \multicolumn{2}{c}{[\si{\micro\meter}]} & \multicolumn{2}{c}{[\si{\milli\meter\per\second}]} &  & \si{\second} \\
    	\midrule
    	\midrule
    	Molasses & 750&/750 & 30.9&/30.9  & \num{4e8} & 2 \\
    	\midrule
    	ODT (\SI{41}{\micro\kelvin}) & 65&/6.2 & 62.7&/44.5 & \num{2.3e7} & 2.7 \\
    	ODT (\SI{41}{\micro\kelvin}) lensed & 86&/7.6 & 51.5&/40.8 & \num{2.2e7} & 2.7 \\
    	\midrule
    	ODT (\SI{4}{\micro\kelvin}) & 15.2&/12.2 & 19.6&/12 & \num{2.7e6} & 4.7 \\
    	ODT (\SI{4}{\micro\kelvin}) lensed & 49&/12.2 & 9.2&/12.9 & \num{2.6e6} & 4.7 \\
    	\midrule
    	BEC & 3.8&/3.3 & 2&/2 & \num{4.3e5} & 8.2 \\
    	BEC lensed & 16.9&/3.3 & 0.55&/2.2 & \num{4.2e5} & 8.2 \\
    	\midrule
    	Advanced & 5&/5 & 2&/2 & \num{1e6} & 1 \\
    	Advanced lensed & 46.1&/46.1 & 0.14&/0.14 & \num{1e6} & 1 \\
    	\bottomrule
    	\bottomrule
    \end{tabular}
    
    \label{tab:instability}
\end{table}
We chose three parameter sets from the here presented measurements of two thermal ensembles released from the optical dipole trap (ODT) with starting temperatures of $T_0 = $~\SIlist{41; 4}{\micro\kelvin} and the BEC.
Besides that we simulated the performance of the interferometer operated with a molasses cooled ensemble combined with a velocity selective Raman pulse of \SI{30}{\micro\second}~\cite{Kasevich91PRL2}, based on typical parameters in our experiment, and an advanced scenario.
For this we assume a BEC with \SI{1e6}{atoms} after a preparation time $t_\text{P}=$~\SI{1}{\second} with a starting expansion velocity of \SI{2}{\milli\meter\per\second}, as anticipated for the VLBAI setup~\cite{Hartwig15NJP, Schlippert2020CPTproceed}.
We extrapolate the performance of our matter-wave lens for this experiment, resulting in expansion velocities of \SI{0.135}{\milli\meter\per\second} corresponding to an equivalent 3D temperature of \SI{200}{\pico\kelvin}.
\section*{Data availability}
The data used in this manuscript are available from the corresponding author upon reasonable request.
\bibliography{main}
\section*{Acknowledgements}
This work is funded by the German Space Agency (DLR) with funds provided by the Federal Ministry of Economic Affairs and Energy (BMWi) due to an enactment of the German Bundestag under Grant Nos. DLR 50WM1641 (PRIMUS-III), DLR 50WM2041 (PRIMUS-IV), DLR 50WM1861 (CAL), DLR 50WM2060 (CARI-OQA), and DLR 50RK1957 (QGYRO).
We acknowledge financial support from the Deutsche Forschungsgemeinschaft (DFG, German Research Foundation)–Project-ID 274200144–SFB 1227 DQ-mat within the projects A05, B07, and B09, and –Project-ID 434617780–SFB 1464 TerraQ within the projects A02 and A03 and Germany's Excellence Strategy—EXC-2123 QuantumFrontiers—Project-ID 390837967 and from “Niedersächsisches Vorab” through the “Quantum- and Nano-Metrology (QUANOMET)” initiative within the Project QT3.
A.H. and D.S. acknowledge support by the Federal Ministry of Education and Research (BMBF) through the funding program Photonics Research Germany under contract number 13N14875. 
\section*{Author contributions}
W.E., E.M.R., and D.S. designed the experimental setup and the dipole trapping laser system.
H.A., A.H., A.R., and D.S. contributed to the design, operation, and maintenance of the laser system and the overall setup.
R.C., E.C. and N.G. set the theoretical framework of this work.
H.A., R.C., C.S., and D.S. drafted the initial manuscript.
H.A., and R.C. performed the analysis of the data presented in this manuscript.
H.A., and R.C. under lead of N.G. and C.S. performed the instability study.
All authors discussed and evaluated the results and contributed to, reviewed, and approved of the manuscript.
\section*{Competing interests}
All authors declare no competing interests.
\end{document}